# Amplification and scintillation properties of oxygen-rich gas mixtures for optical-TPC applications

**L. Weissman**[a*]**, M. Gai**[ab]**, A. Breskin**[c]**, R. Chechik**[c]**, V. Dangendorf**[d]**,
K. Tittelmeier**[d] **and H.R. Weller**[e]

[a] *Laboratory for Nuclear Science at Avery Point, University of Connecticut,
1084 Shennecossett Rd, Groton, CT 06340, U.S.A.*

[b] *Department of Physics, WNSL, Yale University,
New Haven, CT 06520-8124 U.S.A.*

[c] *Dept. of Particle Physics, The Weizmann Institute of Science,
76100 Rehovot, Israel*

[d] *Physikalisch-Technische Bundesanstalt,
38116 Braunschweig, Germany.*

[e] *Department of Physics, Triangle Universities Nuclear Laboratory, Duke University,
Durham, NC 27708, U.S.A.*
*E-mail*: `leonid.weissman@uconn.edu`

ABSTRACT: We studied electron amplification and light emission from avalanches in oxygen-containing gas mixtures. The mixtures investigated in this work included, among others, $CO_2$ and $N_2O$ mixed with Triethylamine (TEA) or $N_2$. Double-Step Parallel Gap (DSPG) multipliers and THick Gas Electron Multipliers (THGEM) were investigated. High light yields were measured from $CO_2+N_2$ and $CO_2+TEA$, though with different emission spectra. We observed the characteristic wave-length emission of $N_2$ and of TEA and used a polymer wave-length shifter to convert TEA UV-light into the visible spectrum. The results of these measurements indicate the applicability of optical recording of ionizing tracks in a TPC target-detector designed to study the cross-sections of the $^{16}O(\gamma,\alpha)^{12}C$ reaction, a central problem in nuclear astrophysics.

[*] Corresponding author.

**Contents**

**1. Introduction**

      This work was motivated by the quest for an oxygen-rich gas mixture for the Optically read out Time Projection Chamber (O-TPC) "target-detector", designed to study the $^{16}O(\gamma,\alpha)^{12}C$ reaction in the $\gamma$ energy range of 7.9–10.0 MeV [1]. Its time-revised $^{12}C(\alpha,\gamma)^{16}O$ reaction is a fundamental unknown in stellar evolution theory and our experiment will enable determining its cross-section in the alpha energy range of below 1.2 MeV.

      The concept of TPCs was developed in the middle of the 1970s [2] and has been broadly used in particle and nuclear physics [3]. In most applications radiation-induced charged-particle tracks are recorded electronically. Optical avalanche readout systems, on the other hand, present an attractive alternative for low counting-rate applications and for the detection of complex event patterns [4]. The first developments of Optical readout TPC (O-TPC) were: 3D deposited-charge imaging in a calorimeter prototype [5] and single-photon imaging in a UV-sensitive gaseous detector for Cherenkov Ring Imaging (RICH) [6]. The optical readout provides genuine 2D (or 3D) images at a moderate cost. It is decoupled from the detector electrodes and vessel, having the advantages of avoiding electronic pickup noise and of high-voltage decoupling from the readout electrodes. Fast optical sensors (e.g. photomultipliers) can provide trigger signals to the slower optical recording system, usually consisting of a gated Intensified CCD camera (ICCD) [5],[7],[8]. The optically read out TPC end-cups are usually simple, and consist of cascaded parallel-mesh avalanche multipliers, within which the light emission is higher compared to that in wire structures [4]. Although the O-TPC concept is well established, only a few applications have been published: in the fields of dosimetry [8],[9] radiography [10], thermal neutron instrumentation [11], as well as in the experiment for the study of two-proton radioactive decay of the $^{45}Fe$ nucleus [12]. In [9]-[11] the authors made use of the GEM (Gas Electron Multiplier [13]) technique.

      A major challenge in designing of an O-TPC detector is choosing of the gas mixture. In addition to the standard requirements of the electron transport parameters for best track quality,

the detector gas should also exhibit good light emission (scintillation) properties to provide efficient and accurate optical imaging. Moreover, in cases where the gas serves also as the reaction target, its composition should match the particular application. A common practice is to use admixtures of Thriethylamine (TEA) or Tetrakis[dimethylamino]ethylene (TMAE) vapors, which usually added to Ar/hydrocarbon mixtures [4]. These vapors are known for their efficient scintillation properties and were found to yield large avalanche-induced light outputs - up to one photon per avalanche electron [14]. Low photoionization thresholds of TEA and TMAE allowed the use of them as "gaseous photocathodes" in UV-photon detectors for Ring Imaging Cherenkov (RICH) devices, including optically read out detectors [6]. However, the use of these vapors often leads to significant complications in detectors and gas-handling systems; e.g. TMAE is very chemically reactive, which causes aging of the detector components. Other gas mixtures, which are simpler to use, were investigated. For example $CF_4$ was found to emit copious light [15] and has been recently used in O-TPCs [9],[12],[16]. In an earlier study $Ar+N_2(2\%)$ was reported to have comparable light yield to that of $Ar+TEA(2\%)$ [17]. Therefore the properties of nitrogen-containing gas mixtures might be of interest for optically read out detectors.

Our goal is the application of an O-TPC detector for the measurement of cross-section of the $^{16}O(\gamma,\alpha)^{12}C$ nuclear reaction [1] at a new generation of gamma-beam facilities such as HI$\gamma$S [18]. An O-TPC setup is well suited for such studies where the detector, with its counting-gas also serving as interaction medium for the gamma-beam, is used to measure the signature of very rare reaction products. Via phase-space considerations these data can be converted to provide the cross-section of the time-reversed $^{12}C(\alpha,\gamma)^{16}O$ nuclear reaction. The advantages of the measurement of the inverse reaction are the significantly larger cross-section and the high luminosity of the available gamma-beam. Furthermore, the reaction products, alpha-particles and carbon ions, are expected to be measured and identified with close to 100% efficiency and with reliable background rejection. The experiment is also expected to yield the angular distribution of the reaction products, and thus a partial wave analysis of the reaction cross-sections [1]. As the cross-section value is expected to be very low, the O-TPC gas-mixture, which constitutes the target volume, should be as dense as possible and contain the maximum possible fraction of oxygen.

Several oxygen-rich gas mixtures were studied in this work, with the objective of reaching both high electron multiplication and high avalanche-induced light yields. We investigated parallel-mesh gaseous multipliers, as well as the novel THick Gas Electron Multipliers (THGEM) [19],[20] as candidates for the O-TPC end-cup. The possibility of using wave-length shifters (WLS) to shift the emitted UV light into the more practical visible spectral range [12],[17] also was considered.

## 2. Experimental setup

The following parameters of different gas mixtures were examined : 1. the total number of emitted photons ("light"); 2. the spectra of emitted light; 3. the total number of avalanche-induced electrons ("charge"); 4. the energy resolution of the detector; 5. the performance of the electron multiplier structure (grid gap structures and THGEM); and 6. the quality of the track images.

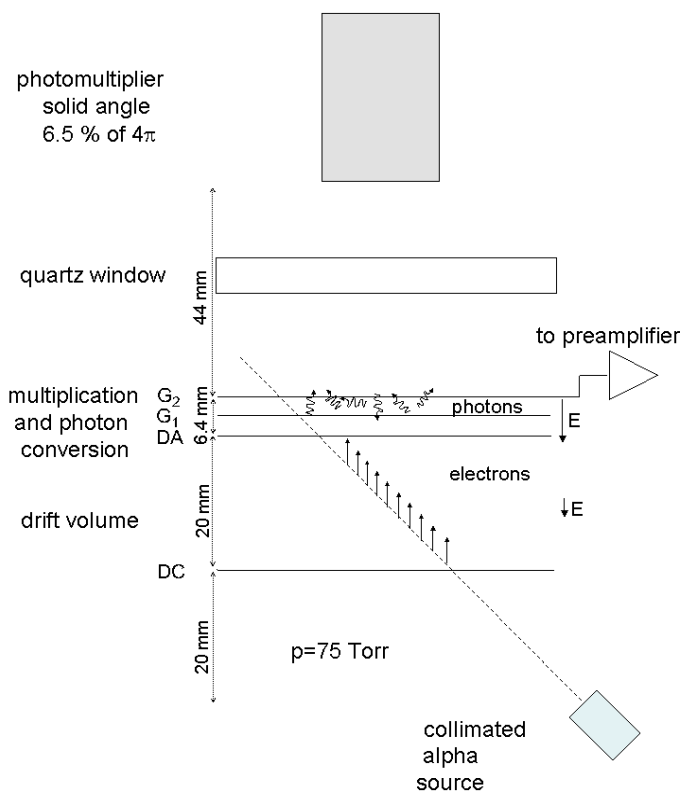

**Figure 1.** Schematic diagram of the O-TPC equipped with a Double-Stage Parallel Gap (DSPG) multiplier, for measurements of charge and light yields in various gases.

A schematic diagram of the O-TPC detector used to measure gain and light emission is shown in figure 1. The detector was filled with gas at a typical pressure of p=75 Torr. This choice of pressure is dictated by the track lengths of the reaction products required for an unambiguous identification of the $^{16}O(\gamma,\alpha)^{12}C$ reaction. Different gas pressures were also used for studying imaging properties of the optical read out system. A drift region that defines the fiducial volume of the detector was created by two 80% optically transparent wire-mesh electrodes separated by 20 mm. The grids were made of stainless steel wires of 50 µm diameter spaced 0.5 mm apart. The grids defining the drift volume are denoted in figure 1 as drift cathode (DC) and drift anode (DA). Typical values of the reduced drift field is 6.5 V/(Torr·cm). Collimated alpha-particles were emitted from a non-spectroscopic gold-coated $^{241}$Am radioactive source (mean energy of ~4.5 MeV and FWHM of energy distribution of 350 keV). The source was placed below the drift region, irradiating the detector at 45 degrees relative to the drift field lines, as shown in figure 1. The range of these 4.5 MeV alpha-particles, at a pressure of 75 Torr, is 10-22 cm, depending on which gas mixture was tested. The primary electrons created in the sensitive volume along the alpha-particle track drift towards the drift anode (DA) and experience two-step multiplication in the Double-Stage Parallel Gap (DSPG) multiplier. The multiplier was defined by the three mesh planes DA, $G_1$ and $G_2$, placed at a 3.2 mm distance from each other (figure 1). In some other configurations studied here, the electrons were multiplied in a sequence of a THGEM multiplier and a parallel-grid gap. The avalanche multiplication process yields a proportional amount of photons due to excitation of the gas molecules, thus most of the photon emission occurs at the final amplification stage

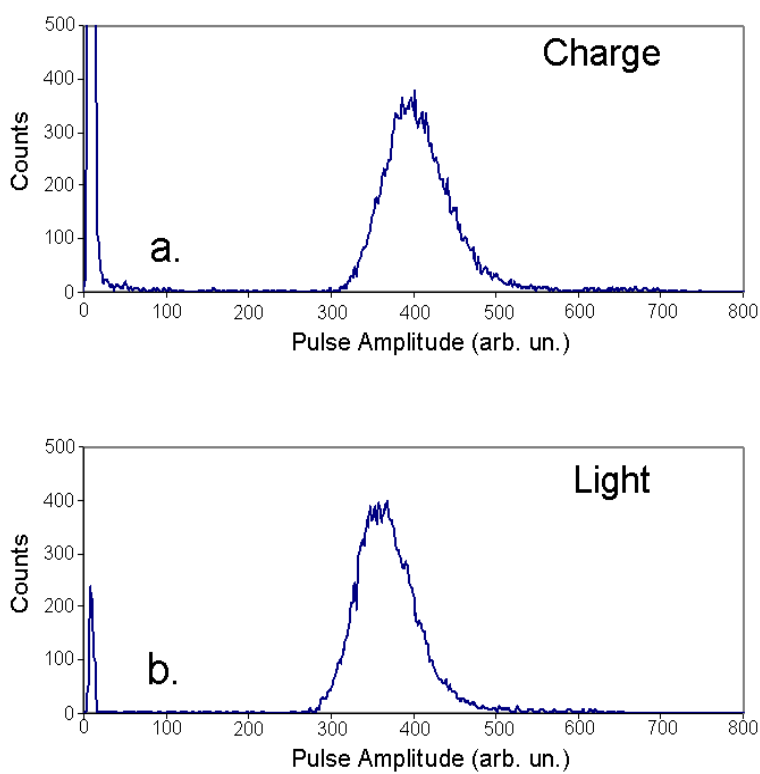

**Figure 2.** Typical collected charge (a) and light (b) spectra emitted by a $CO_2+N_2$ (10%) gas mixture. The broad peaks (FWHM~20%) are due to the large distribution of the energy loss of alpha-particles emanating at different angles and having different track length, in the sensitive volume.

[8][14]. The amounts of avalanche-induced charge and light strongly depend on the values of the electrical fields in the multiplication gaps. The reduced electric field in the amplification gaps was varied between 10 to 65 V/(Torr·cm) and light emission was typically observed at reduced-field values higher than 20 V/(Torr·cm). The charge produced by the avalanche was collected on $G_2$ and was amplified by a calibrated charge-sensitive preamplifier and a linear spectroscopy amplifier. The signals were analyzed using a PC-based multichannel analyzer (Amptek, MCA8000A). The emitted photons were detected by a UV-sensitive Photonis XP-2020Q photomultiplier (PMT). The PMT was placed 44 mm from the light emission region, behind a quartz view port. It collected the avalanche-induced light within a solid angle of 6.5% of $4\pi$. The signal from the PMT anode was amplified by a linear spectroscopy amplifier and analyzed using the same MCA.

A second data acquisition (DAQ) system was also used for analyzing the charge and light spectra. This CAMAC-based DAQ (KMAX by Sparrow Inc.) incorporated an eight-channel ORTEC ADC, AD811. The use of this DAQ system allowed one to establish 2D charge versus light correlation plots.

Typical avalanche-induced charge (electron yield distribution) and light (photoelectron yield distribution) spectra are shown in figure 2. Note that the observed FWHM (of ~ 20%) of the distributions was due to dispersion in the energy loss of alpha-particles. This was partially due to the initial alpha-particle energy distribution, but mostly because of the geometrical variations of the track lengths. These variations were associated with imperfect collimation of the source. More detailed studies of the energy resolution of the detector were performed with a spectroscopic-quality source (see below).

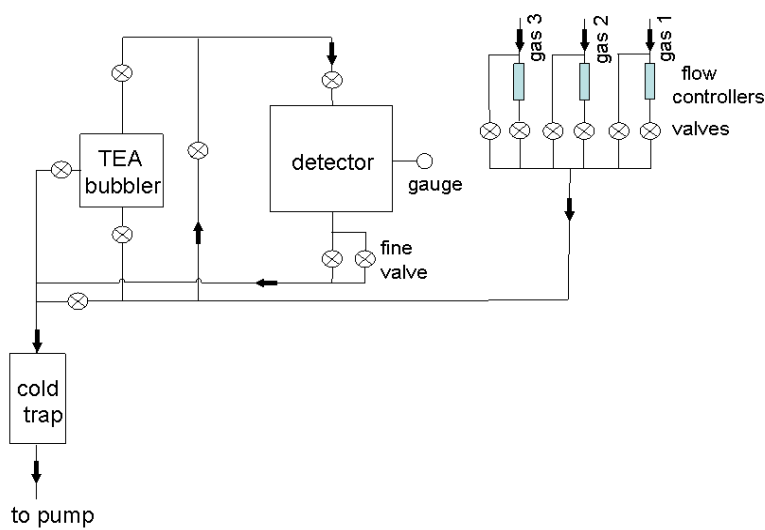

**Figure 3.** Schematic diagram of the gas handling system.

The electron charge amplification was determined from the ratio of the measured charge (i.e. the number of avalanche electrons), to the calculated number of primary electrons produced in the drift region. The latter was calculated for each gas mixture using the stopping power tables [21] and the known W-values – the mean energies of formation of an electron-ion pair. The total number of photons emitted into $4\pi$ was calculated from the number of photoelectrons measured at the PMT photocathode multiplied by its wave-length dependent quantum efficiency, as provided by the manufacturer, and the calculated solid-angle of the PMT. The number of photoelectrons produced in the PMT photocathode was determined from the PMT pulse-height spectrum (figure 2) normalized to the single-photoelectron peak. This peak was obtained from the thermal-emission spectrum at high PMT bias voltages of 2.1-2.3 kV and was corrected to the PMT gain difference at the operating voltages of 1.0-1.4 kV. The measured PMT multiplication curve was used for this correction. The uncertainties in the PMT calibration procedure and in the quantum efficiency of the PMT cathode (up to 30% according to the manufacturer) are the main factors contributing into the overall systematic uncertainty in the determination of absolute light yield. This overall uncertainty amounted to up to 50%. The number of photons per electron, ph/e, in the avalanche was readily deduced from measured charge and light.

Transmission filters were placed in front of the PMT in the measurements of the light emission spectra from the mixtures of interest (see below). A set of transmission filters with cutting edges at the wavelengths ($\lambda_s$) of 280, 290, 305, 320, 335, 345, 360, 375, 400, 420, 435, 455, 475, 485, 530, 580 and 630 nm was used for this purpose.

High-purity (99.999%) gases were used in all measurements. The detector was operated in continuous gas-flow mode using the gas handling system shown in figure 3. The system allows for mixing up to three different gases. The partial pressures of the gas components were set by mass-flow controllers and the overall pressure in the detector was controlled by a capacitive pressure gauge. All parameters of the gas system were set by a MKS146 controller (MKS Instruments Inc.). TEA vapors could be added by bubbling the mixed gas through a reservoir filled with TEA liquid. A cold trap mounted on the pump (figure 3) eliminated TEA vapor contamination.

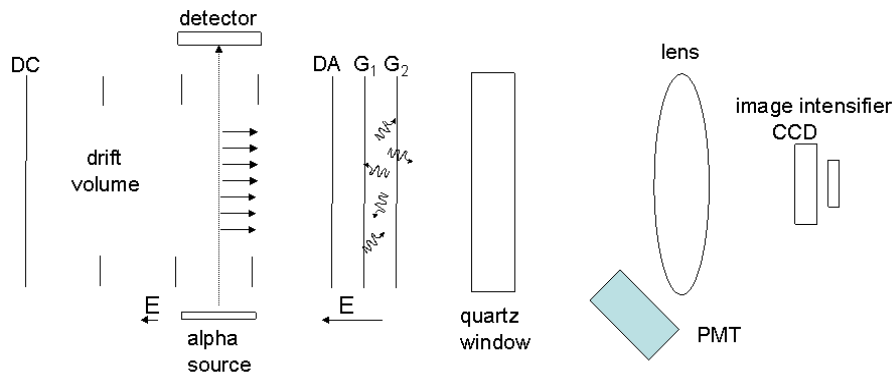

**Figure 4**. Schematic diagram of the O-TPC used for recording the tracks of alpha-particles.

The investigated oxygen-rich gas mixtures were based on carbon dioxide ($CO_2$) and nitrous oxide ($N_2O$ or "laughing gas"). Neither $CO_2$ nor $N_2O$ exhibit light emission compatible with optical readout systems sensitive to the UV-to-visible range. Thus other efficient gaseous light emitters, like TEA or TMAE, must be added. The TEA light emission spectrum is centered around 280 nm [22] and requires the use of expensive UV optics. TMAE emits visible light peaking at 480 nm [23], which is compatible with standard optical elements. However, TMAE has high chemical reactivity and low vapor pressure [14], requiring heating of the detector and gas system to rather high temperatures in order to reach a light output comparable to TEA. Therefore only the addition of TEA vapors was practiced in this work. The partial pressure of TEA vapor was controlled by varying the temperature of the liquid in the bubbler. The dependence of TEA vapor pressure on temperature (T) in the applicable temperature range is described by the empirical equation:

$$p = 52.0 \times e^{(4100 \times (1/293 - 1/T(K)))} \text{ Torr [24]} \qquad (1)$$

A different experimental setup was used for recording images of alpha-particle tracks. The detector and the optical readout system are shown in figure 4. The electrodes of the imaging O-TPC had a similar geometry to the one presented in figure 1, thus its performance is expected to be the same. A non collimated $^{241}$Am source was placed to the side of the drift volume.

A small parallel-plate proportional detector (PPAD) was placed opposite of the alpha-particle source, providing the event trigger for the optical read out chain. Thus, only alpha-particles traversing the whole sensitive volume were registered. Three guard rings biased by a resistive chain were placed between the drift anode and cathode, thus ensuring the homogeneity of the drift field. The ionization electrons produced along the alpha-particle tracks drifted towards the DA. The avalanche-induced light produced in the last multiplication stage was detected by the optical read out system. A quartz UV lens with a focal length of 150 mm and an aperture of F = 2.5 focused the light onto the photocathode of an Image Intensifier (Proxifier BV 2562) coupled to a CCD camera [8]. The optical efficiency of the system was estimated to be $4 \times 10^{-4}$, and the quantum efficiency of the photocathode was 0.2. This yielded an effective photon detection efficiency of $8 \times 10^{-5}$. The image intensifier was gated by a high voltage (HV) pulse, which was initiated by the PPAD alpha-particle trigger. The CCD analog output signal was digitized by a 12 bit PCI digitizer board (Spectrum MI 3025). After careful CCD background subtraction, track images with a 10 bit useful dynamic range were obtained.

## 3. Experimental procedures and results

### 3.1 Measurements of "charge" and "light" yields

During the initial search for the optimal gas, we performed numerous short tests with various gas mixtures using the setups shown in figure 1 and figure 4. The tested gases and gas mixtures included pure $CO_2$, pure $N_2O$, pure $O_2$, $CO_2$+TEA, $N_2O$+TEA, $CO_2$+$N_2$, $N_2O$+$N_2$, $CO_2$+$CF_4$, $CO_2$+isobutane and $CO_2$+$O_2$. Gas mixtures containing $O_2$ did not exhibit sufficient gain and did not allow for stable operation of the detector. Pure $CO_2$ and $N_2O$ gases did not exhibit sufficient light yields in the UV-to-visible spectral range. In these preliminary tests the best results were obtained with $CO_2$ and $N_2O$ mixed with TEA or $N_2$. Hence, the detailed

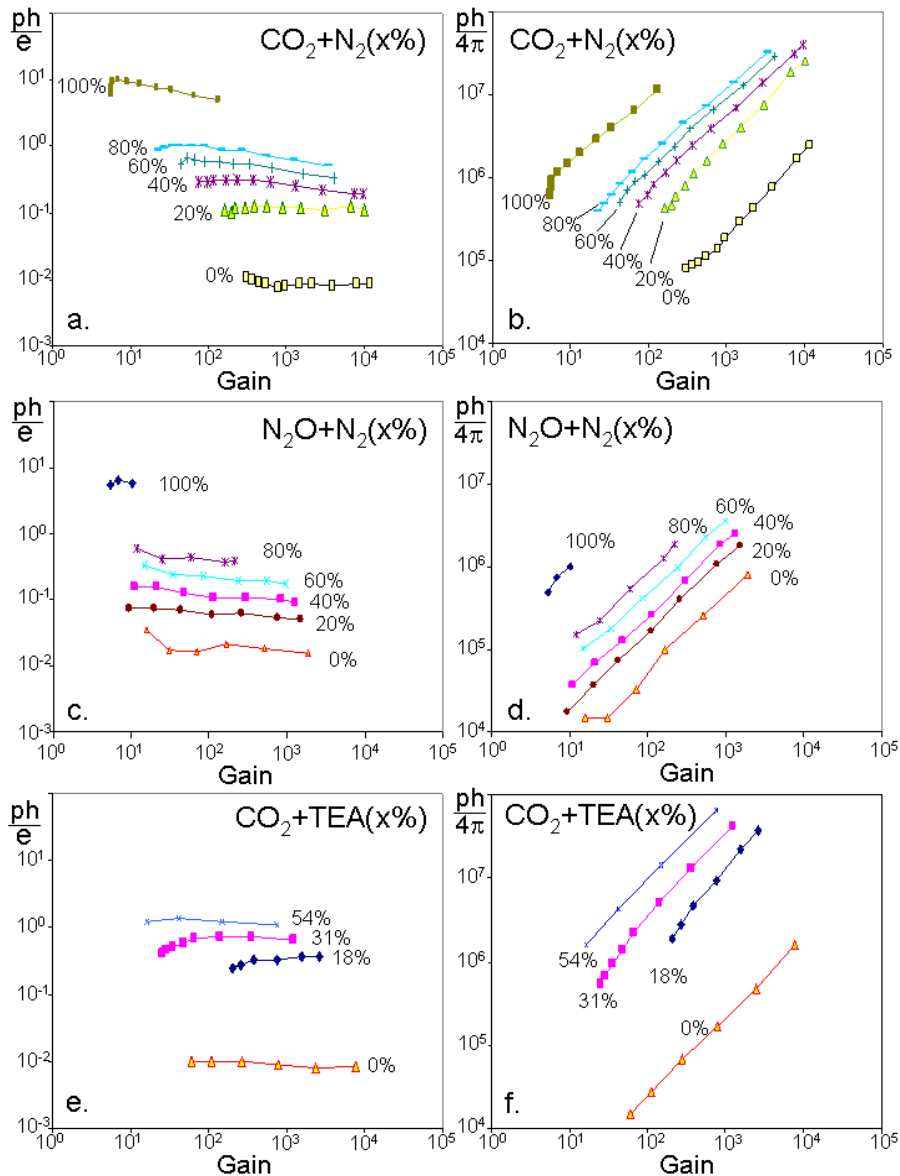

**Figure 5.** Electron and photon yields for three gas mixtures, $CO_2$+$N_2$ (a,b), $N_2O$+$N_2$ (c,d) and $CO_2$+TEA (e,f). The total number of emitted photons and the ratio of photons to electrons in the avalanche are plotted as functions of the charge gain. The $N_2$ and TEA concentrations in the mixtures are shown.

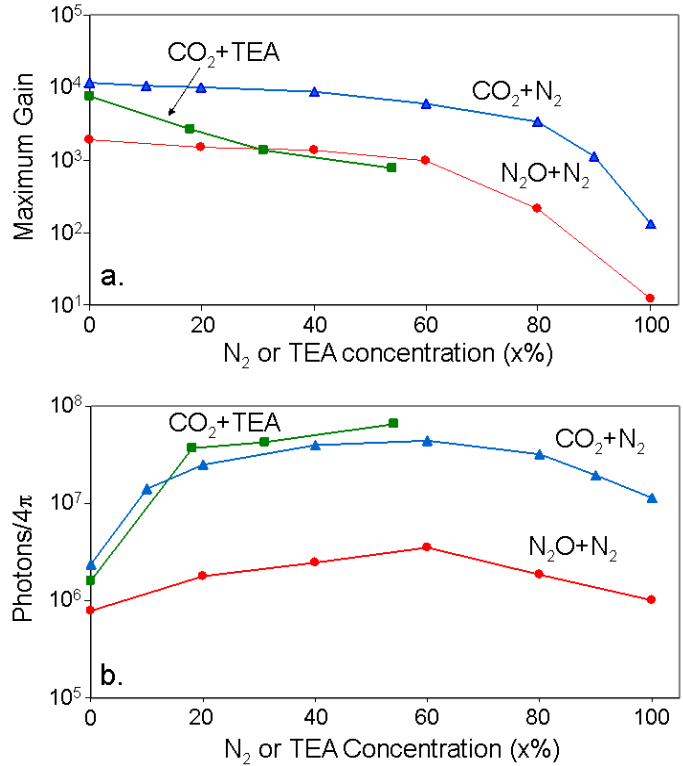

**Figure 6.** The maximum gain (a) and the maximum number (b) of emitted photons as a function of concentration of $N_2$ or TEA. The data presented for $CO_2+N_2$, $N_2O+N_2$ and $CO_2+TEA$.

systematic measurements were performed only for $CO_2+N_2$, $N_2O+N_2$ and $CO_2+TEA$ gas mixtures.

Electron multiplication and light emission were measured while varying the electric fields in the two multiplying gaps as described in the previous section (figure 1). During the measurements the electric field in the first gap (between grids DA and $G_1$) was kept at the highest possible value below discharge, while the field in the second gap (between grids $G_1$ and $G_2$) was varied. The results of the measurements are presented in figure 5. The concentrations of $N_2$ and TEA in the mixtures are indicated in the figure. The TEA concentrations corresponded to the TEA bubbler temperatures of -4, 5 and 15 °C. The summary of the maximum attainable gains and numbers of emitted photons as a function of $N_2$ or TEA concentrations is provided in figure 6.

As shown in figure 5, the ph/e ratio does not decrease by less than a factor of 2 with the increase of the multiplication gain by two orders of magnitude. Hence, the amount of light is essentially proportional to the charge multiplication. This behavior is different from the one usually observed in noble-gas mixtures, where the ph/e ratio decreases by an order of magnitude over a similar multiplication range [12],[17]. This phenomenon was explained by a decrease of the excitation cross-sections and an increase of the ionization probability with increasing field values [17]. One can observe from figure 5 that the photon-to-electron ratios are similar for the $CO_2+N_2$ and $N_2O+N_2$ mixtures with the same $N_2$ concentrations. However, the use of $CO_2+N_2$ is more favorable since it allows for the application of higher electric fields, resulting in higher multiplication factors and higher light. Taking into account the higher oxygen content of $CO_2$, it is clear that $CO_2+N_2$ is more suitable for the proposed experiment [1] than $N_2O+N_2$. Although the ph/e ratio for $CO_2$+$N_2$ mixtures are somewhat lower than for the corresponding $CO_2$+TEA mixtures, the charge multiplication factors are higher for the first. Therefore, the maximum number of emitted photons are comparable for these two cases as shown in figure 6.

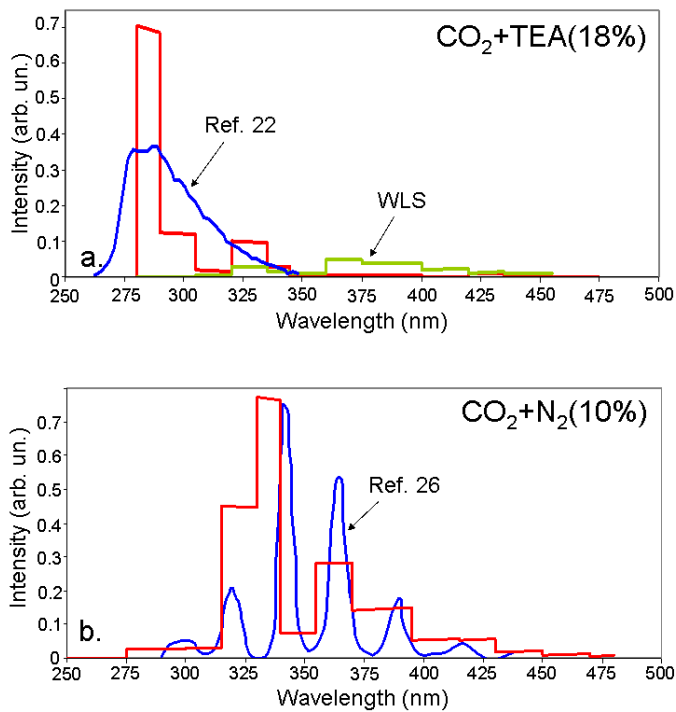

**Figure 7.** (a) Emission spectra of $CO_2$+TEA(18%) and (b) $CO_2$+$N_2$ (10%)(b) mixtures. The measured emission spectra of TEA and $N_2$ [22,26] are shown for comparison. The spectrum of $CO_2$+TEA(18%) measured with an unknown polymer wave-length shifter is also shown.

The ph/e ratio for $CO_2$+$N_2$ increases linearly with $N_2$ concentration. However, the maximum charge multiplication factor decreases with $N_2$ concentration, most likely because of the lower breakdown voltages of $N_2$ [25]. Therefore, an increase of the $N_2$ concentration beyond 50% leads to a decrease of the maximum number of emitted photons (figure 6b). In fact, the relatively low concentrations of $N_2$, (e.g. 10%) results in the maximum photon yields of approximately $10^7$ photons per alpha-particle track. This is very well suited to the objective of maximizing the oxygen content in the gas mixture for the planned experiment [1].

**3.2 Measurement of light emission spectra**

The light emission properties of the two best oxygen-containing mixtures, $CO_2$+$N_2$ and $CO_2$+TEA, were investigated. Light-emission spectra of $CO_2$+$N_2$ (10%) and $CO_2$+TEA(18%) mixtures were measured by utilizing the set of transmission filters (see above). Light distribution spectra similar to those shown in figure 2b were collected for each of the filters placed in front of the PMT. All other conditions (voltages, pressure, gas mixture composition) were kept identical. The measured pulse-height centroids are proportional to the number of transmitted photons with wavelengths higher than the cutting edge of the corresponding filter, $\lambda > \lambda_s$. The difference between the photon yields measured in successive exposures with different filters allowed us to derive the emission spectra. The quantum efficiency of the PMT bialkali photocathode, which is almost constant within the investigated spectral range, was also taken into account. The deduced emission spectra are shown in figure 7.

– 9 –

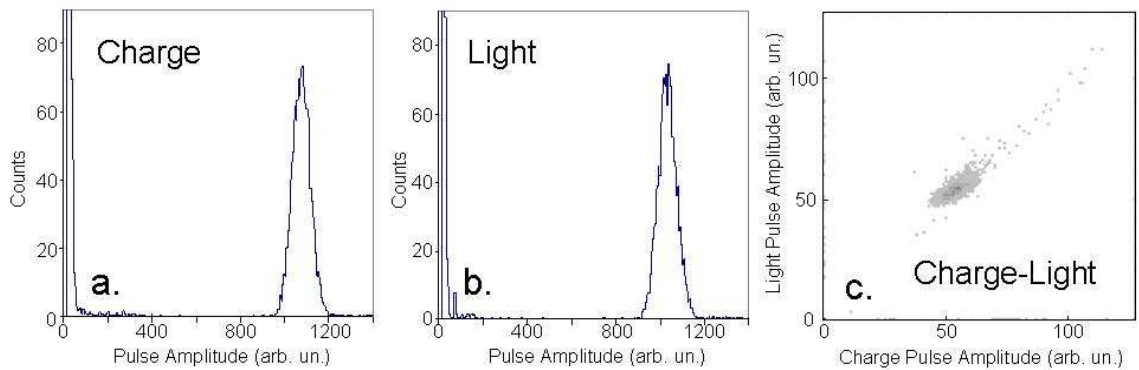

**Figure 8.** The charge (a) and light (b) distributions obtained with the double-collimated spectroscopic 5.58 MeV alpha-source. The alpha particles were traversing the middle part of the sensitive volume, parallel to the grid electrodes. The light-to-charge correlation plot of the same data is shown in (c). The dark area corresponds to the FWHM of the peak.

As pointed out in [22], the emission spectra of noble gases mixed with TEA are dominated by the TEA emission independently of the noble-gas used. The results of our measurements show that the $CO_2$+TEA mixture has similar spectral characteristics with dominant peak at 280 nm (figure 7a). Asymmetric emission bands, of up to 350 nm, were attributed to the radiative deexcitation of the excited TEA molecules [22]. In addition, we repeated the measurements using a polymer Wave Length Shifter (WLS) of the same type as used in [12]. The WLS foil was made of polystyrene doped with dyes. The spectrum measured with the WLS is shown in figure 7a. The WLS shifted the TEA light in the spectral range centered at about 380 nm; however considerable photon losses were incurred.

The light emitted from avalanches in the $CO_2$+$N_2$ (10%) mixture is shown in figure 7b. Most of the emitted light corresponds to the 337 nm $N_2$ line. However, a significant amount of light is emitted at longer wave lengths (377 and 391 nm lines of $N_2$) which approach the visible spectral range. The spectrum of nitrogen fluorescence induced by alpha-particle excitation of pure nitrogen gas [26] is shown for comparison in figure 7b. The light emission of $N_2$ is in the soft UV near the visible spectral range. This light emission is beneficial for most applications when compared to the TEA light emission, making possible the use of standard optics.

**3.3 Measurement of energy resolution**

The energy resolution of the O-TPC detector is of importance for rejecting anticipated background events in the future experimentation [1]. The data shown in figure 2 were obtained without optimization of the energy resolution. In these measurements, variations of the alpha-particle paths within the drift and multiplication regions led to a large spread in the deposited energy. On the other hand, the spread of the energy of alpha-particles emerging from the gold-plated source was founded to be 8% (FWHM).

In order to measure the energy resolution of the O-TPC, we used a spectroscopic $^{241}$Am source ($E_\alpha$ = 5.49 MeV). A 3 mm diameter collimator was placed next to the source and a second, 1 mm diameter, collimator was placed at a 25 mm distance from the source. The source was mounted to the side of the drift volume, and the collimated alpha-particles impinged perpendicular to the drift field. The CAMAC-based DAQ system (see above) was used in these measurements.

Typical charge and light spectra and a two-dimensional correlation spectrum are shown in figure 8. The spectra were taken in $CO_2$+$N_2$(10%) at 75 Torr. The voltages on DC, DA, $G_1$

– 10 –

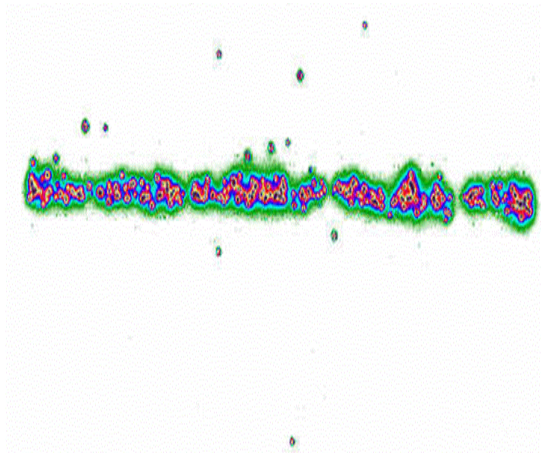

**Figure 9.** A typical image of alpha-particle tracks taken with a $CO_2+N_2$ (10%) mixture at 75 Torr. The track length is 90 mm.

and $G_2$ electrodes were -2.2, -1.2, 0 and 1.75 kV respectively. These data exhibit good energy resolution: 8.3% (FWHM) for the charge spectrum and 9.0% (FWHM) for the light spectrum. The light-to-charge correlation spectrum presented in figure 7c demonstrates the proportionality between charge and lights. The energy resolution of the O-TPC proposed for the $^{16}O(\gamma,\alpha)^{12}C$ experiment is expected to be better due to complete stopping of alpha-particles and carbon ions in a larger O-TPC volume. For example, an energy resolution of 2% was obtained with an O-TPC in [16] where the alpha-particles were stopped within the drift volume of the detector.

### 3.4 Optical imaging of alpha-particle tracks

Images of the alpha-particle tracks were acquired with the imaging O-TPC viewed by an intensified CCD camera, in a setup described in detail in [8] and in the previous section (figure 4).

A typical image of an alpha-particle track taken in a $CO_2+N_2$ (10%) mixture at 75 Torr is shown in figure 9. The energy loss of the alpha-particles in the sensitive volume is approximately 1.0 MeV. The visible part of this track was 90 mm long. The applied voltages on the DC, DA, $G_1$ and $G_2$ electrodes were -1900 V, -800 V, +900 V and +2000 V, respectively. This image was taken at the relatively low multiplication value of approximately $10^3$, corresponding to $4.3 \times 10^6$ photons emitted in $4\pi$ and 340 photoelectrons produced in the photocathode of the image intensifier. The latter number was obtained by multiplying the number of photons emitted in $4\pi$ by the total detection efficiency of the optical system ($8 \times 10^{-5}$, see above). As seen in figure 6 the detector gain and the total number of avalanche-emitted photons can be increased by an order of magnitude for the $CO_2+N_2$ (10%) mixture.

Image properties of $CO_2$+TEA mixtures were also investigated. For example, track images were optically recorded in 27 Torr of $CO_2$+TEA. The liquid TEA was kept at -5 °C, yielding a $CO_2$+TEA(50%) mixture. Unlike in $CO_2+N_2$(10%), we observed "halo" photons around the alpha-particle tracks, as shown figure 10a. Images taken with prompt (400 ns) and delayed HV gate pulses to the image intensifier are shown in figure 10b and figure 10c, correspondingly. The delayed emission of the halo photons clearly indicates that they were emitted by a secondary process. It suggests a photon feedback mechanism, where photons from the primary avalanche were reabsorbed in the drift region and induced further ionization electrons. Since re-absorption occurred within a certain volume around the primary avalanche,

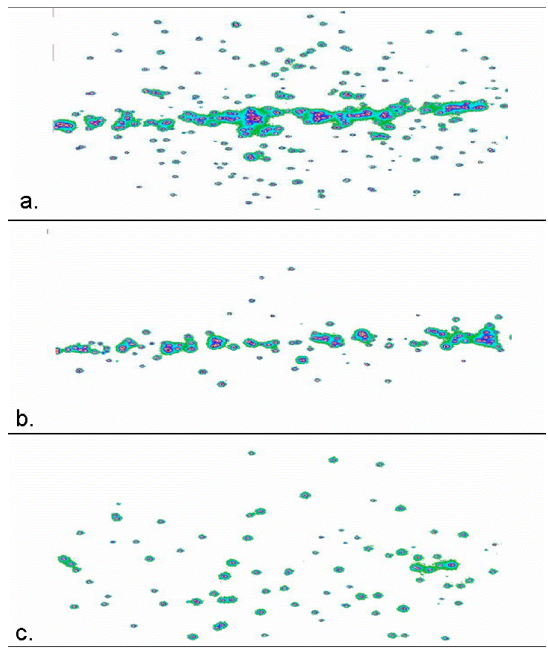

**Figure 10.** (a) A typical image of an alpha-particle track with $CO_2$+TEA(50%) mixture recorded with an ungated image-intensifier. (b) Similar image taken with the image intensifier bias gated by a 400 ns prompt pulse, and (c) a delayed gate on the image intensifier

these secondary electrons led to secondary avalanches that were scattered around the primary ones. Note that such a phenomenon was not observed for the $CO_2+N_2$ (10%) gas mixture or for the pure TEA vapor [8]. Therefore, the observed photon feedback is not associated with photo-effects on the electrodes or on the walls, but is a specific feature of the $CO_2$ + TEA mixtures.

### 3.5 Measurements with Thick GEM multipliers

Following the successful implementation of O-TPC detectors with GEM-based light-emitting multipliers [10],[11],[16] we investigated the light emission properties of the simpler THick Gas Electron Multipliers (THGEM) [19],[20],[27],[28]. In these hole-multipliers, the avalanche is confined to the holes resulting in photon-feedback suppression. Two series of 10 cm diameter THGEMs (Print Electronics [29]), referred below as THGEM A and THGEM B, were produced from double-sided copper clad G-10 circuit board of thicknesses 0.4 mm (A) and 1.6 mm (B) respectively. Other THGEM dimensions are presented in table 1 and a photograph of a typical THGEM used in this work is shown in figure 11. The copper was etched around the hole's rim to reduce the probability of discharges at high voltages [19],[20]. The leakage currents across the THGEMs were measured in ambient air as a function of the applied voltage. Those yielding the lowest leakage current (~ 1 nA at 1 kV) were chosen for the tests.

|  | THGEM A | THGEM B |
|---|---|---|
| Thickness (mm) | 0.4 | 1.6 |
| Drilled hole diam. (mm) | 0.5 | 1.0 |
| Etched hole in Cu diam. (mm) | 0.7 | 1.2 |
| Hole pitch (mm) | 1.0 | 1.5 |
| Area (cm$^2$) | 78.5 | 78.5 |
| Hole diam./Thickness | 1.25 | 0.625 |

**Table 1.** Dimension of THGEM used in the tests.

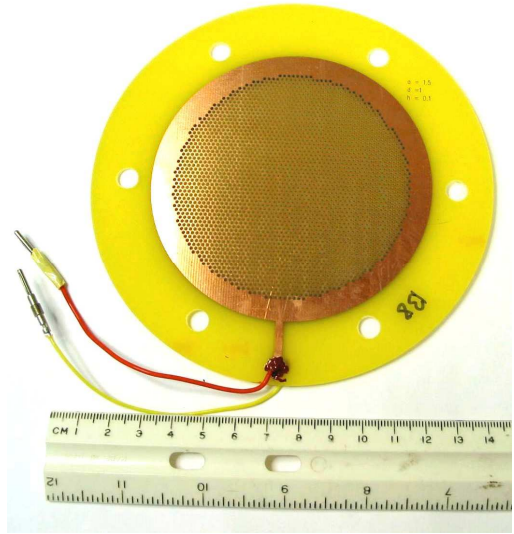

**Figure 11.** A photograph of a THGEM used in the measurements.

The THGEM was installed in the O-TPC detector replacing the DSPG multiplier. One side of the THGEM was kept at ground potential and was used as a drift anode, as shown in the insert of figure 12. A grid electrode, $G_1$, was placed 3.2 mm downstream from the THGEM as an optional second amplification and scintillation stage. The amplified charge could be collected on the THGEM or on $G_1$. The detector was operated in $CO_2+N_2$ (10%) mixture at 75 Torr. The reduced electric field in the drift volume was similar to the field in the previous measurements (6.5 V/(cm·Torr)). In the first test the charge was collected on the THGEM anode while the voltage on the $G_1$ electrode was kept at ground potential, providing a strong reverse field and full charge collection. The multiplication gains as a function of the applied voltage measured with THGEM A and B are shown in figure 12. For the maximum applicable voltages, prior to discharge across the THGEM, we measured gains of $1\text{-}1.5\times10^3$. Cascading two THGEM elements separated by a 1 cm gap did not increase the total gain beyond $3\times10^3$ at any possible combination of the THGEM voltages. This suggests that the space charge limit of the gain was reached at least in one hole of the second THGEM at the present conditions. The somewhat lower gains of the THGEM compared to the ones obtained with the DSPG configuration is the subject of further investigation.

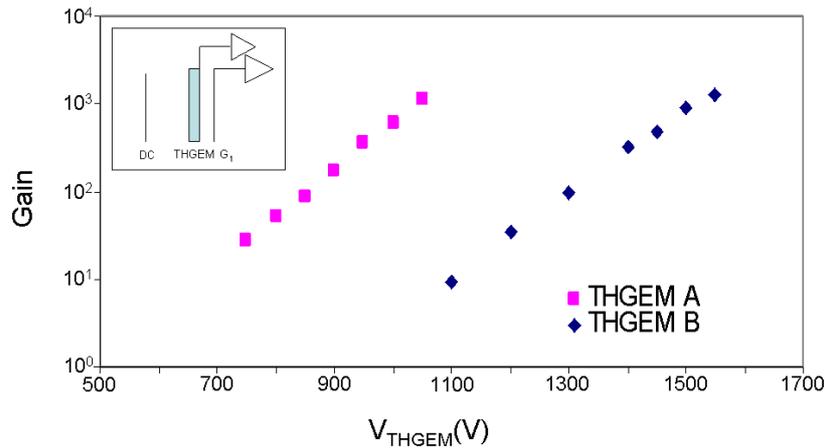

**Figure 12.** Dependence of the gain of two different THGEMs (see table 1) on the applied voltage. The measurements were done in a $CO_2+N_2$ (10%) at 75 Torr. A schematic diagram of the electrode configuration is shown in the insert. In these measurements the charge was collected directly on the THGEM's anode.

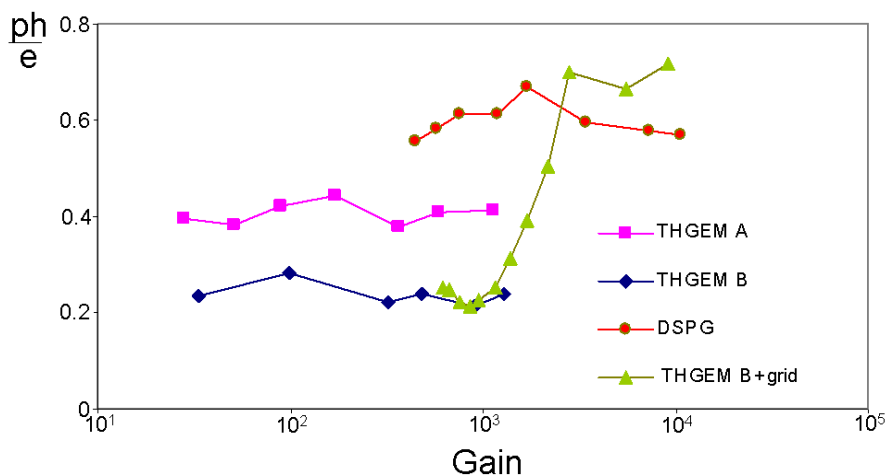

**Figure 13.** Photon-to-electron ratio measured with THGEM A, THGEM B (see table 1), the DSPG and with THGEM B followed by a grid electrode. All measurements were done for a $CO_2+N_2(10\%)$ mixture at 75 Torr.

The dependence of ph/e ratios as a function of the multiplication gain is shown in figure 13. As in the case of DSPG multiplier, the ratios are almost independent of the voltage across the THGEM. However, they are consistently lower than those obtained with two-grid multipliers for the same $CO_2+N_2$ (10%) gas mixture (see figure 13). In addition, the ph/e ratios observed for the thicker THGEM B are almost a factor of two lower than those obtained from THGEM A. This suggests either geometrical blocking of the light within the holes, or implies that the scintillation process is less intense in the confined volume. The ph/e ratio in the modified configuration, with an additional amplification process taking place in the gap between the THGEM and the $G_1$, is also shown in figure 13. In this mode the voltage across the THGEM B was fixed at the maximum value (~1.4 kV) and the amplification field in the THGEM-$G_1$ gap varied while the avalanche charge was collected on the $G_1$ grid. In this configuration a higher ph/e ratio was measured at the highest applicable voltages (figure 13).

## 4. Summary and discussion

We studied the properties of oxygen-rich gas mixtures. The gases investigated primarily involved $CO_2+N_2$, $N_2O+N_2$ and $CO_2+TEA$ mixtures. $CO_2+N_2$ and $N_2O+N_2$ gas mixtures were found to exhibit good light-emission yields, when compared to $CO_2+TEA$ mixtures. While the ph/e ratios of $CO_2+N_2$ and $N_2O+N_2$ mixtures similar, the maximum amplification gain and total light yield significantly higher in the case of $CO_2+N_2$. In addition, $CO_2+N_2$ contains twice as much oxygen. The maximum number of photons emitted in the corresponding $CO_2+N_2$ and $CO_2+TEA$ mixtures comparable. Thus, the observed high light emission of nitrogen-containing mixtures confirms the results reported almost two decades ago [17].

– 14 –

Light emission by $CO_2$+TEA at a wave length of 280 nm requires the use of rather expensive UV optics or the use of wave length shifters. However, to follow the WLS option in the proposed O-TPC, a more elaborate investigation is required. The wavelength of the light emitted by the $CO_2$+$N_2$ is in the soft UV, near visible range. This is more favorable for design of an economical optical system.

The energy resolution of the O-TPC detector with the $CO_2$+$N_2$(10%) mixture was found to be 8-9% (FWHM). In planned experiment the energy resolution of the O-TPC is expected to be better due to complete stopping of alpha-particles and carbon ions in the drift volume. The obtained energy resolution is sufficient to remove the anticipated background events in the proposed $^{16}O(\gamma,\alpha)^{12}C$ experiment.

Good quality images of alpha-particle tracks were obtained in the $CO_2$+$N_2$(10%) mixture using the prototype O-TPC detector. These images did not exhibit the photon feedback effect, obtained from the $CO_2$+TEA mixture. In the proposed experiment [1] we expect the alpha-particles originating from the $^{16}O(\gamma,\alpha)^{12}C$ reaction to lose between 1.0 and 1.8 MeV in the drift region of the O-TPC. The anticipated optical system will have a total detection efficiency of ~$1.4 \times 10^{-4}$, which is higher than the efficiency of the system used in this work. This estimation is based on: 1. $30 \times 30$ cm$^2$ active area of the TPC detector (object size); 2. an image intensifier with a cathode diameter of 40 mm (image size); 3. quantum efficiency of the image intensifier of 0.2; and 4. a lens with aperture of F = 1.1 and focal length of 150 mm. In the planned experiment the energy deposited by the photodisintegration products will be similar to the energy loss of alpha-particles in the present tests. Compared to the conditions at which the image in figure 9 was taken the gain and light outputs can be increased by the order of magnitude. All these facts ensure that a similar or even better quality of tracks could be achieved in the experiment.

Thus, it can be concluded from our studies that $CO_2$+$N_2$(10%) is a well suited gas mixture for the planned experiment. The gas mixtures containing nitrogen have many advantages over the standard gas mixtures with photosensitive vapors and may find broader applications in the gas detectors field.

Further research is needed for evaluating the best light-emitting electron multiplier. The accent should be on maximal light yield, energy resolution, lack of secondary avalanches (halo) and track quality. The highest gains of the order of $10^4$ were achieved with the DSPG and THGEM plus grid configurations. The somewhat lower charge and light yields obtained with THGEMs will be address in the future.

**Acknowledgements**

The authors acknowledge the help of Mr. M. Klin and Mrs. A. Raanan of the Weizmann Institute, for preparing the gas system and for performing preliminary light-emission measurements. The imaging O-TPC and its read out and data analysis system as well as the detector setup for the light and charge measurements, were made by Dr. G. Laczko, formerly at PTB, Braunschweig. The research was funded in part by TUNL at Duke University, the US department of energy grant number DE-FG02-94ER40870, the Benoziyo Center for High-Energy Research at the Weizmann Institute, the Israel Science Foundation - project 151/01 and by the Yale-Weizmann Collaboration, ACWIS, New-York. L. Weissman is grateful to Mr. T. Kading for revision of the manuscript.

A. Breskin is the W. P. Reuther Professor of Research in the Peaceful use of Atomic Energy.